\documentstyle[amsfonts,aps,prb,twocolumn]{revtex}
\begin{document}
\draft
\title{Vector magnetic hysteresis of hard superconductors}

\author{A. Bad\'{\i}a$^{1}$ and C. L\'opez$^{2}$}
\address{$^{1}$Departamento de F\'{\i}sica de la Materia Condensada-I.C.M.A., 
\\$^{2}$Departamento de Matem\'atica Aplicada,
\\C.P.S.U.Z., Mar\'{\i}a de Luna 3, E-50.015
Zaragoza (Spain)}
\date{\today}
\maketitle
\vspace{5mm}
\begin{abstract}

Critical state problems which incorporate more than one component for 
the magnetization vector of hard superconductors are investigated. 
The theory is based on the minimization of a cost functional ${\cal 
C}[\vec{H}(\vec{x})]$ which weighs the changes of the magnetic field 
vector within the sample. We show that Bean's simplest prescription 
of choosing the correct sign for the critical current density $J_c$ 
in one dimensional problems is just a particular case of finding the 
components of the vector $\vec{J}_c$. $\vec{J}_c$ is determined by 
minimizing ${\cal C}$ under the constraint $\vec{J}\in\Delta 
(\vec{H},\vec{x})$, with $\Delta$ a bounded set. Upon the selection 
of different sets $\Delta$ we discuss existing crossed field 
measurements and predict new observable features. It is shown that a 
complex behavior in the magnetization curves may be controlled by a 
single external parameter, i.e.: the maximum value of the applied 
magnetic field $H_m$.

\end{abstract}
\vspace{10mm}
\narrowtext
%
%
\section{Introduction}
The irreversible magnetization of type-II superconductors was 
successfully treated by Bean's critical state model (CSM) in the 
early sixties.\cite{bean} The basic idea is that the material 
develops a maximum (critical) current density in those regions which 
have been affected by an electric field, opposing to the magnetic 
flux changes. With remarkable simplicity, the CSM could explain the 
dependence of the magnetization on the macroscopic dimensions of the 
sample, as well as the observed hysteresis. Both features are a 
manifestation of the non-equilibrium thermodynamic processes which 
take place in the experiments and, thus, cannot be explained by the 
Abrikosov's flux line lattice (FLL) theory.\cite{abrikosov} 
Furthermore, hard type-II materials develop such a pronounced 
hysteresis that the reversible contribution from the equilibrium FLL 
may be usually neglected. Nowadays, the hit of Bean's intuition is 
well understood in terms of the FLL dynamics in the presence of 
pinning centers.\cite{richardson} In the process of punching vortices 
in or out the superconductor one produces metastable equilibrium 
states for which the gradient in the density of vortices is maximum. 
This results in the development of the macroscopic critical current 
density $J_c$ and corresponds to the balance between a repulsive 
vortex-vortex interaction and attractive forces towards the pinning 
centers.

Bean's {\em ansatz} has permitted the development of the theory with 
a maximum simplicity while incorporating the main experimental facts. 
However, it was early recognized insufficient for problems in which 
lattices of non-parallel flux tubes must be considered. For such 
cases, the current density within the sample is no more a vector with 
fixed direction, and additional prescriptions are required. On the 
other hand, a wealth of phenomena have been reported along the last 
decades associated to crossed field measurements. 
Refs.\onlinecite{leblanc,park,leblanc2,voloshin} provide some 
outstanding experiments, and the interested reader is directed to the 
papers cited therein for more complete landscape of the topic. 
Roughly speaking, when the experimental scenario is such that 
compression and rotation of vortices are induced, both the scalar 
repulsion between rigid parallel vortices and the cutting energy 
barrier for adjacent twisted flux lines come into play. In 
particular, this has consequences on the transport properties of 
type-II superconductors in a magnetic field, as the pinning of the 
FLL is strongly influenced by its internal stiffness.\cite{sudbo}

Several efforts have been made for the interpretation of rotating 
field experiments. Bean himself developed an approach for such 
problems.\cite{beanjap} However, to our knowledge, the most extensive 
general critical state theory is due to Clem and 
P\'erez-Gonz\'alez\cite{clemgonzalez1,clemgonzalez2,clemgonzalez3,clemgonzalez4,clemgonzalez5,clemgonzalez6,clemgonzalez7,clemgonzalez8} 
who have developed the so-called double critical state model (DCSM). 
These authors have provided a physical basis for the limitations 
on the current flow both parallel and perpendicular to the local 
magnetic induction. Thus, the so-called parallel critical current 
density $J_{c\parallel}$ and perpendicular critical current density 
$J_{c\perp}$ were introduced. $J_{c\parallel}$ relates to the flux cutting 
threshold, whereas $J_{c\perp}$ stands for the {\em conventional} 
depinning current density. Recently, a more sophisticated approach was introduced, the so-called {\em two-velocity hydrodynamic model}.\cite{voloshin} Essentially, this theory incorporates the flux pinning and cutting phenomena within the framework of FLL's which consist of two vortex subsystems. Under certain circumstances it is fully equivalent to the DCSM, but, different, non-expected scenarios are also predicted.

In earlier work\cite{badia,badia2} we have envisioned the CSM realm as a phenomenological approach which can be formulated by means of a 
variational problem with constraints. This permits to deal with a 
wide class of restrictions on the current density. Thus, the optimal 
control (OC) theory\cite{oc} allows the minimization of a cost 
functional ${\cal C}[\vec{H}(\vec{x})]$ which weighs the changes of 
the magnetic field vector within the sample under the constraint 
$\vec{J}\in\Delta (\vec{H},\vec{x})$, with $\Delta$ a bounded set. 
This general formulation gives freedom to fix the particular region 
$\Delta$ of feasible current densities. For instance, the isotropic 
hypothesis $|\vec{J}|\leq J_c$ corresponds to choosing $\Delta$ as a 
disk, and the DCSM conditions $|J_{\perp}|\leq J_{c\perp}$, 
$|J_{\parallel}|\leq J_{c\parallel}$, to an oriented rectangle.

In this work we present a detailed discussion of the variational 
principle (Sec.\ref{varpri}), as well as the general aspects about 
the Hamiltonian equations which arise from our formulation (Sec. 
\ref{hameq}). Sec.\ref{mag} gathers a series of simulations of the 
field penetration profiles for two different selections of the 
control space $\Delta$. In particular, it is explicitly shown that 
the DCSM becomes a particular case of our treatment. Then, Sec. 
\ref{magloops} is devoted to observe some properties of the magnetization loops which can be derived from the simulated profiles. In Sec.\ref{conclusions} we give a general overview of the model and discuss the compatibility of different choices of $\Delta$ with available experimental data.
%
%
\section{Variational principle}
\label{varpri}
In this section we will show that the coarse-grained electrodynamics 
of type-II superconductors in the critical state may be 
formulated as a minimization problem. From a more conventional side, 
on using the DCSM approach, one can obtain the successive field 
penetration profiles in a magnetization process by means of Maxwell 
differential equations, and the constitutive $\vec{E}(\vec{J})$ law 
for the material. These equations, together with the boundary 
conditions allow to solve for the dependencies 
$\vec{H}(\vec{x},t),\vec{J}(\vec{x},t),\vec{E}(\vec{x},t)$. As it is 
usual in hard superconductivity we are assuming 
$\vec{B}=\mu_{0}\vec{H}$.

Now we recall that many physical theories, which are typically posed 
by a differential equation statement have been interpreted as the 
minimization of a given functional. For instance, the Lagrangian 
formulation of classical mechanics relies on the equivalence of 
Newton laws and the stationarity condition for the action $S=\int L\, 
dt=\int (T-V)\, dt$, when the time evolution of the system is 
determined. Below, we derive the functional ${\cal C}$ whose 
minimization is equivalent to the standard approach for the critical 
state in superconductors. In order to gain physical insight, we will 
infer the CSM equations after considering some aspects of the more 
familiar eddy-current problem in normal metals ($\vec{E}=\rho 
\vec{J}$). Assuming a discretization scheme in which $\vec{H}_{\rm 
n}$ stands for the magnetic field intensity at the time layer 
$n\delta t$ and Amp\`ere's law ($\nabla\times\vec{H}=\vec{J}$), the 
successive field profiles in a magnetic diffusion process may be 
obtained by the finite-difference expression of Faraday's law
\begin{equation}
\label{eqn1}
\mu_{0}\frac{(\vec{H}_{\rm n+1}-\vec{H}_{\rm n})}{\delta t}=
-\nabla\times\vec{E}=
-\rho \nabla\times(\nabla\times \vec{H}_{\rm n+1}) \; ,
\end{equation}
which defines a differential equation for $\vec{H}_{\rm n+1}$ in the 
spatial degrees of freedom. Then $\vec{H}_{\rm n+1}$ can be solved in 
terms of the previous field distribution $\vec{H}_{\rm n}$ and the 
boundary conditions at the time layer $(n+1)\delta t$. The 
minimization statement can be found by {\em inversion} of the 
Euler-Lagrange equations with respect to some Lagrange density that 
belongs to Eq.(\ref{eqn1}). The first term is straightforwardly 
inverted by integration with respect to the field $\vec{H}_{\rm 
n+1}$. This gives $\mu_{0}(H_{\rm n+1}^{2}/2-\vec{H}_{\rm 
n}\cdot\vec{H}_{\rm n+1})/\delta t$, i.e.:
\[
\frac{\partial (H_{\rm n+1}^2/2-\vec{H}_{\rm n}\cdot\vec{H}_{\rm 
n+1})}{\partial \vec{H}_{\rm n+1}}=
\vec{H}_{\rm n+1}-\vec{H}_{\rm n}\; .
\]
The second term is a bit more involved as it cannot be transformed to 
an exact derivative with respect to the fields. However, if one introduces the notation $(\nabla\times\vec{H}_{\rm 
n+1})_i=\epsilon_{ijk}\partial_{x_j}{H}_{\rm n+1,k}$  (with 
$\epsilon_{ijk}$ the Levi-Civita tensor) it can be identified from
\begin{eqnarray}
\partial_{x_j}\frac
{\partial(\epsilon_{klm}\partial_{x_l}{H}_{\rm n+1,m}
\,\epsilon_{kpq}\partial_{x_p}{H}_{\rm n+1,q})}
{\partial (\partial_{x_j}{H}_{\rm n+1,i})}=
\nonumber
\\
-2[\nabla\times(\nabla\times\vec{H}_{\rm n+1})]_i \; .
\nonumber
\end{eqnarray}
Thus, Eq.(\ref{eqn1}) can be  rewritten as:
\begin{equation}
\label{eqn2}
\frac{\partial{\cal F}_{\rm n+1}}{\partial H_{\rm n+1,i}}-
\partial_{x_j}
\frac{\partial{\cal F}_{\rm n+1}}{\partial (\partial_{x_j}{H}_{\rm n+1,i})}
=0
\; ,
\end{equation}
where
\begin{eqnarray}
{\cal F}_{\rm n+1}=&&
\mu_{0}(H_{\rm n+1}^2/2-\vec{H}_{\rm n}\cdot\vec{H}_{\rm n+1})+
\nonumber
\\
&&\rho\delta t(\nabla\times\vec{H}_{\rm n+1})\cdot 
(\nabla\times\vec{H}_{\rm n+1})/2\; .
\nonumber
\end{eqnarray}
From variational calculus it is known that Eq.(\ref{eqn2}) 
corresponds to the necessary first order conditions for the 
functional ${\cal C}_M=\int_{\Omega}\! {\cal F}_{\rm n+1}$ to have a 
minimum with respect to the field $\vec{H}_{\rm n+1}$ 
(Euler-Lagrange equations).\cite{arfken}

Finally, it is apparent that for minimization purposes ${\cal C}_M$ 
may be completed by means of a constant term to the form
\begin{eqnarray}
{\cal C}_{M}=&&
\frac{\mu_{0}}{2}\int_{\Omega}\!|\vec{H}_{\rm n+1}-\vec{H}_{\rm n}|^{2}+
\nonumber
\\
&&\frac{1}{2}\int_{\Omega}\!\rho\delta t (\nabla\times\vec{H}_{\rm n+1})\cdot 
(\nabla\times\vec{H}_{\rm n+1})\; ,
\label{eqn3}
\end{eqnarray}
as $\vec{H}_{\rm n}(\vec{x})$ is considered to be given.

In conclusion, from the mathematical point of view, minimizing {\em 
C}$_M$ for each advancing time layer is just an equivalent statement 
of Faraday's laws in time discretized form. In addition, 
Eq.(\ref{eqn3}) allows a physical interpretation which can be used 
for comparison and generalization to other systems. Notice that the 
minimization of ${\cal C}_{M}$ balances the {\em screening} term 
$|\vec{H}_{\rm n+1}-\vec{H}_{\rm n}|^{2}$ and the isothermal {\em 
entropy production} term $(\dot{\cal S}=\vec{E}\cdot\vec{J}/T)$. This 
can be identified as a quite general property of dynamical systems 
subjected to dissipative forces. The quasistationary time evolution of the system holds a compensation between some {\em inertia} term and the 
irreversible loss of energy. The minimum principle for the global 
entropy production rate lies behind the previous statement. It was 
introduced by I. Prigogine\cite{prigogine} in the context of linear 
irreversible thermodynamics and applies for non-equilibrium 
stationary states. As a simple exercise, the reader can 
check that minimizing $m|\vec{v}_{\rm n+1}-\vec{v}_{\rm 
n}|^{2}+\gamma \vec{v}_{\rm n+1}\cdot\vec{v}_{\rm n+1} \delta t$ with 
respect to $\vec{v}_{\rm n+1}$ one reproduces the discretized statement of Newton's law for the motion of a particle against a viscous damping 
force. In the case of eddy currents $|\vec{H}_{\rm n+1}-\vec{H}_{\rm 
n}|^{2}$ stands for the magnetic field inertia, including both the 
magnetostatic energy term $\vec{H}_{\rm n+1}^{2}$ and the term 
$\vec{H}_{\rm n}\cdot\vec{H}_{\rm n+1}$, which describes the work 
against electromotive forces.

Eventually, we notice that hard type-II superconductors may be 
treated by a modification of the functional {\em C}$_M$ for normal 
metals. We want to emphasize that the concept of hard material is 
associated to a limiting case for the $\vec{E}(\vec{J})$ 
characteristic, i.e: the electric field is zero for current densities 
below a critical value ($J_c$) and abruptly raises to arbitrarily 
large values if $J_c$ is overrun. In fact, this vertical graph limit 
is attained for the experimental situations in which the excitation 
typical period is large as compared to the magnetic diffusion time 
constant $\tau_{\rm f} \sim \mu_{0}L^{2}/\rho_{\rm f}$, where 
$\rho_{\rm f}$ stands for the flux flow resistivity and $L$ is some 
typical length of the sample. Then, for current densities below 
$J_c$, $\dot{\cal S}$ vanishes as no electric field is generated in 
stationary conditions. On the other side, in the approximation of 
arbitrarily large flux flow resistivity, $\dot{\cal S}$ would diverge 
if $J>J_c$ and, thus, ${\cal F}_{\rm n+1}$ must be minimized 
constraining the current density to $J\leq J_c$. We wish to remark 
that the definition of critical current density may be done in a very 
general sense. In fact, one can postulate vertical $\vec{E}(\vec{J})$ 
relations for definite directions of space as it is the case of the 
DCSM, in which one uses parallel and perpendicular projections with 
respect to the local magnetic field. One could even dictate that huge 
dissipation occurs whenever the current density vector $\vec{J}$ lies 
outside some allowed region $\Delta$, generating an almost instantaneous change of the magnetic profile.

In the light of the previous discussion, the evolutionary critical 
state profiles can be obtained either by using Maxwell equations and 
a vertical $\vec{E}(\vec{J})$ law or the principle:

{\em In a type-II superconducting sample $\Omega$ with an initial
field profile $\vec{H}_{\rm n}(\vec{x})$, and under a small change of 
the external drive, the new profile $\vec{H}_{\rm n+1}$ minimizes the 
functional}
\begin{equation}
\label{eqnvarpri}
{\cal C}[\vec{H}_{\rm n+1}(\vec{x})]
=\frac{1}{2}\int_{\Omega}\! | \vec{H}_{\rm n+1} -
\vec{H}_{\rm n} |^{2} \; ,
\end{equation}

{\em with the boundary conditions imposed by the external source, and 
the constraint $\nabla\times\vec{H}_{\rm n+1}\in\Delta (\vec{H}_{\rm 
n+1},\vec{x})$.}

In this work we will concentrate on two particular cases which can be 
justified in terms of the underlying physical mechanisms. First, we 
will analyze the isotropic case $|\vec{J}|\leq J_c$ (i.e.: $\Delta$ 
is a disk). Then the variational formulation of the DCSM $(J_{\parallel}\leq 
J_{c\parallel},J_{\perp}\leq J_{c\perp})$ will be given. In this case 
$\Delta$ is an oriented rectangle.

The next section is devoted to some aspects on the mathematical formalism of the optimal control theory. We will show that this method provides a very convenient framework for applying the variational principle stated above.

%
%
\section{Hamiltonian equations}
\label{hameq}
From the mathematical point of view, the problem of minimizing the
action integral ${\cal C}[\vec{H}_{\rm n+1}(\vec{x})]$ 
[Eq.(\ref{eqnvarpri})] with ${\vec H}_n({\vec x})$ given, and ${\vec 
H}_{n+1}$ fulfilling
the differential equation
\begin{equation}
\nabla \times {\vec H}_{n+1} = {\vec J} \in \Delta
\subset {\Bbb R}^3
\nonumber
\end{equation}
is a problem of variational calculus with nonholonomic constraints 
(constraints on  the derivatives of the state variables). Being the 
constraint set $\Delta$ a compact region with boundary, we must take 
into account the possibility that the minimum is either reached in an interior point or in a boundary point. The machinery generalizing the classical variational calculus to sets with boundaries is the maximum principle of optimal control, introduced by Pontryagin.\cite{oc} OC is a standard theory in 
engineering, where dynamical systems are actuated by external 
controls (limited to some maximal values) in order to get a desired behaviour. 
The given system of differential equations, including the effect of 
external actions is named the control system, and the integrand of 
the minimizing functional is the cost or performance function. The 
mathematical problem here is similar, with Ampere's law as control 
system  and performance function  ${{1}\over {2}}| {\vec H}_{n+1} - 
{\vec H}_n|^2$, although of course the bounded variable $\vec J$ is 
not an external control. As we will see later on, the algebraic 
condition of maximality embraces the cases of interior and boundary 
optimal points, therefore including the classical equations of 
variational calculus for unconstrained problems, and the modified 
equations for minimal solutions in the boundary. In addition to the 
original reference,\cite{oc} the interested reader is invited to 
review Refs.\onlinecite{leit,knowl} for a more comprehensive and 
topical statement of the OC machinery.

In this paper, in order to simplify the presentation, we fix the 
geometry of the sample to be an infinite slab, so that the control 
system (Ampere's law) becomes by symmetry considerations a simple 
system of ordinary differential equations. Other geometries, as the 
cylinder, give way to different control systems, and the  case of a 
finite sample generates a more involved system of partial 
differential equations. For the slab geometry, with applied magnetic 
field parallel to the faces we have
\[
{{d{\vec H}_{n+1}}\over {dx}} = {\vec f}({\vec H}_{n+1}, {\vec u},x) \quad
{\vec f} \in \Delta
\]
as control system. Hereafter, we take the X axis perpendicular to the 
slab faces and the origin of coordinates at the midplane. By construction, the  vector ${\vec f}=(0, f_y,f_z)$ is orthogonal to the physical variable $\vec J$,
$f_y = J_z$, $f_z = - J_y$. Notice that, despite the applied rotation, we use the same notation for the allowed 
control set $\Delta$. Note also that the 
function $\vec f$ has dependence on the local magnetic field (this 
allows to include the usual models $J_c(|{\vec H}|)$, and possible 
anisotropy), on the position (for potentially  inhomogeneous 
materials) and on some independent coordinates $\vec u$, the control 
variables in OC notation, parameterizing the region $\Delta$.

The first step in the theory of constrained variational calculus is the
definition of a Hamiltonian density, containing the performance 
function, and momenta variables that play the role of Lagrange 
multipliers for the constraints.
\begin{equation}
\label{eqnhamiltonian}
{\cal H}( {\vec H}_{n+1}, {\vec u}, {\vec p}, x)\equiv{\vec p}\cdot {\vec
f}-{{1}\over {2}}| {\vec H}_{n+1} - {\vec H}_n(x)|^2 \; .
\end{equation}
Recall that here ${\vec H}_n(x)$ is a given profile. It is important to 
notice that $\cal H$ is not the usual Hamiltonian function of 
Classical Mechanics in phase space; here it depends on the usual 
state ${\vec H}_{n+1}$ and momenta $\vec p$ variables, but 
additionally on the control $\vec u$. Therefore, the associated 
equations are not only the Hamiltonian differential equations but 
also an extra algebraic condition of maximality, in order to 
determine the extra variables $\vec u$. Denoting by ${\vec 
H}_{n+1}^*(x)$, ${\vec p}^{\,*}(x)$ and ${\vec u}^*(x)$ the optimal 
solution functions (i.e. minimizing ${\cal C}$ and satisfying the 
control system), the OC equations are
\begin{equation}
\label{eqnham1}
{{d{\vec H}_{n+1}^*}\over {dx}} =  {{ \partial{\cal H}}\over
{\partial {\vec p}}} = {\vec f}
({\vec H}_{n+1}^*, {\vec u}^{*},x) \; ,
\end{equation}
the adjoint equations for the momenta
\begin{eqnarray}
\label{eqnham2}
&&{{d{\vec p}^{\,*}}\over {dx}} = - {{\partial {\cal H}}\over {\partial
{\vec H}_{n+1}}}=
\nonumber
\\
&&{\vec H}_{n+1}^*-{\vec H}_n(x)
   - {\vec p}^{\,*}\cdot {{\partial {\vec f}}\over {\partial {\vec H}_{n+1}}}
({\vec H}_{n+1}^*, {\vec u}^*, x) \; ,
\end{eqnarray}
and the algebraic condition of maximality
\begin{equation}
\label{eqnmax}
{\cal H} ( {\vec H}, {\vec u}^*, {\vec p}, x) \geq {\cal H} ( {\vec
H}, {\vec u}, {\vec p}, x)
\qquad \forall \;{\vec f}({\vec H}, {\vec u},x) \in \Delta \; .
\end{equation}

${\vec H}$ and $\vec p$ are fixed in this last condition. In 
general, this allows to find a relation ${\vec u}^*({\vec H},{\vec p}, 
x)$. When this relation is replaced in the former Hamiltonian 
equations [Eqs.(\ref{eqnham1}) and (\ref{eqnham2})], a well posed 
system of ordinary differential equations appears.

For the class of Hamiltonian $\cal H$ described above, the algebraic 
condition of maximality is fulfilled for a vector $\vec f$ with 
maximum projection over the momentum $\vec p$. As a simple exercise, 
the reader is invited to check that this rule produces the Bean CSM 
for one dimensional problems (parallel vortices). Thus, on choosing 
${\vec H}=H_{z}\hat{z}$ one has $f=-J_{y}$. The restriction on the 
current density reads $J_{y}=uJ_{c}$ with $|u|\leq 1$. Then, 
Eq.(\ref{eqnmax}) gives $u^{*}={\rm sgn}(p)$ and this leads to $f=\pm 
J_{c}$ in agreement with Bean's prescription. Remarkably, for the 
case of multidimensional systems (non-parallel vortices) the 
maximality condition will give the {\em critical current} vector 
both in modulus and direction. Note that $\vec{J}$ is determined 
dynamically, through the evolution of the momentum vector. The 
specific details on the distribution rule for the components of 
$\vec{J}_{c}$ depend on the control space $\Delta$ and will be 
discussed in the next section.

As regards the boundary conditions that must be used for obtaining 
the solution profiles ${\vec H}_{n+1}^*(x)$ from Eqs.(\ref{eqnham1}) 
and (\ref{eqnham2}), several considerations must be made. Firstly, in 
the absence of demagnetizing effects, ${\vec H}_{n+1}^*$ is 
determined on the faces of the slab by continuity of the external 
applied field. For the remaining boundary conditions, two typical 
situations appear. In the first case, the modified penetrating 
profile ${\vec H}_{n+1}^*(x)$ equals the former profile ${\vec H}_n(x)$ 
before reaching the centre of the slab. This gives place to an 
(unknown in advance) point $x^*$ such that ${\vec H}_{n+1}^*(x) = 
{\vec H}_n(x) \quad \forall\; x \leq x^*$. The free boundary 
condition ${\cal H}(x^*)=0$ applies in this case, and allows to 
determine $x^*$. In the second case, the new profile never meets the 
former one, and the value of ${\vec H}_{n+1}^*(0)$ is unknown; then, 
the corresponding transversality condition ${\vec p}(0) = {\bf 0}$ 
completes then the number of required boundary conditions.
%
%
\section{Field penetration profiles}
\label{mag}
The Hamiltonian formalism developed above can be expeditiously 
applied to calculate the field penetration profiles for magnetization 
processes in which non-parallel vortex configurations are enforced. 
As the applied field is assumed to be the same on both sides of the 
slab under consideration, we can restrict to the interval $0\leq 
x\leq a$. By virtue of the symmetry, the same behavior appears for 
$-a\leq x\leq 0$.
%

\subsection{Isotropic model}
\label{circ}
First, we will derive some results for the field penetration process 
in the so-called isotropic hypothesis: $|\vec{J}|\leq J_{c}$. In 
order to show the capabilities of the model, a field dependence 
$J_{c}(H)$ will be allowed. In particular, the Kim's model\cite{kim} 
expression $J_c(H)=J_{c_{0}}/(1+H/H_{0})$ will be used. Recall that 
the microstructure dependent parameters $J_{c_{0}},H_{0}$ are 
included. For convenience, the following dimensionless units are 
introduced: $x$ is given in units of $a$, $H$ in units of $H_{0}$, 
and $J$ in units of $H_{0}/a$. Then, the statement of Amp\`ere's law, 
together with the critical current restriction read
\[
\frac{d\vec{H}_{\rm n+1}}{dx}=
\vec{f}(\vec{H}_{\rm n+1},\vec{u},x)=
\frac{\beta\vec{u}}{1+|\vec{H}_{\rm n+1}|}
\quad |\vec{u}| \leq 1
\; .
\]
Above we have introduced the dimensionless constant 
$\beta=J_{c_{0}}a/H_{0}$ and the so-called {\em control variable} 
$\vec{u}$, which is a vector within the unit disk $D$. This means 
that the current density belongs to a disk of variable radius, 
depending on the local magnetic field. Thus, following the OC terminology, we 
have the {\em control equations} for the {\em state variables} 
$\vec{H}_{\rm n+1}(x)$.

Next, we require the minimization of the {\em functional}
${\cal C}[\vec{H}_{\rm n+1}(x)]$ constrained by the state equations. 
Following the OC machinery introduced in Sec.\ref{hameq}, the 
maximization of the associated Hamiltonian ${\cal H}$ 
[Eq.(\ref{eqnhamiltonian})] with respect to the control variable 
$\vec{u}$ leads to the condition $\vec{u}^{*}=\vec{p}^{\,*}/p^{*}$. This has the physical counterpart $|\vec{J}| =J_c(H)$, i.e.: within 
the isotropic model, the maximum allowed current density modulus 
$J_c$ is carried within those regions which have been affected by the 
perturbation. Then, the Hamiltonian equations that provide the field 
profile are
\begin{mathletters}
\label{eqncanslab:all}
\begin{eqnarray}
\frac{dH_{\rm 
n+1,i}^{*}}{dx}&=&\frac{p_{i}^{*}}{p^{*}}\,\frac{\beta}{1+H_{\rm 
n+1}^{*}}
\label{eqncanslab:a}
\\[1ex]
\frac{dp_i^{*}}{dx}&=&H_{\rm n+1,i}^{*}-H_{\rm n,i}+
\frac{\beta p^{*}H_{{\rm n+1},i}^{*}}
{H_{\rm n+1}^{*}(1+H_{\rm n+1}^{*})^2} \; .
\label{eqncanslab:b}
\end{eqnarray}
\end{mathletters}
Eventually, appropriate boundary conditions must be supplied to solve 
this set of equations. As it was discussed in Sec.\ref{hameq} these 
conditions are given by the external drive values, together with the 
field penetration regime. If the new profile matches the old one at a 
point $0< x^{*}< 1$ (partial penetration) one uses $\vec{H}_{\rm 
n+1}^{*}(x^{*})=\vec{H}_{\rm n}(x^{*})$ and ${\cal H}(x^{*})=0$. 
Within the full penetration regime, however, one uses the momenta 
transversality conditions $\vec{p}^{\,*}(0)={\bf 0}$.

In order to illustrate the bundle of phenomena which can be expected in 
crossed field measurements, we have calculated the penetration 
profiles for a zero field cooled sample to which a constant 
excitation $H_{zS}$ is applied, followed by cycling stages of the 
other field component at the surface $H_{yS}$. We want to emphasize 
that, in order to approximate the continuum evolution, small 
increments of $H_{yS}$ have been applied in the iterative solution of Eqs.(\ref{eqncanslab:a}) and (\ref{eqncanslab:b}). However, for clarity, only a selection of representative field and current density curves are 
depicted. The value of $\beta$ is set to unity for definiteness hereafter.

%
\subsubsection{Initial magnetization process}
\label{initial}
Fig.\ref{fig1} displays the penetration profiles for the initial 
magnetization process that is induced by increasing the surface field 
$H_{yS}$. We notice that the conventional $H_{z}(x)$ profile (which 
was obtained by the standard CSM) is pushed towards the center as 
$H_{yS}$ increases. In physical terms, a current flow $J_z$ is 
required for shielding the new field component $H_{y}$. Owing to the 
constraint $|\vec{J}| =J_c(H)$, this results in a reduction of $J_y$ 
(the slope of $H_{z}$) and $H_z$ develops a subcritical behavior. 
Just for clarity, we have only plotted advancing fronts until the 
centre of the sample is reached. Ongoing profiles will be shown in 
the next subsections.

%
\subsubsection{Low field hysteresis}
\label{lowfield}
Here, we analyze the hysteresis effects which can be observed when 
the surface field is cycled. First, the maximum applied field 
($-H_m\leq H_{yS}\leq H_m$) is chosen so as to keep a zero field 
value at the midplane. We define the penetration field $H_{m}^{*}$ as 
the value of $H_{yS}$ for which the front reaches the point $x=0$. 
This is somehow equivalent to the so-called partial penetration regime in the standard CSM. However, notice that now $H_{m}^{*}$ depends on the previously set value of $H_{zS}$. Fig.\ref{fig2} displays the 
results. It can be observed that $H_{y}(x)$ basically shows a 
conventional critical state behavior, whereas $H_{z}(x)$ develops a 
prominence. This shape relates to the corresponding slope reduction 
near the surface ($x=1$) and the subsequent reentry in the 
unperturbed profile. Physically, the distribution rule for the 
current density vector enhances the $J_z$ component near the surface 
in order to shield the change in $H_{y}$ (i.e.: reducing $\int 
|\vec{H}_{\rm n+1}-\vec{H}_{\rm n}|^{2}dx$). Then, $J_y$ diminishes 
and, therefore $H_{z}(x)$ flattens. An internal slope increase is 
obviously required for the previous profile $H_{\rm n,z}(x)$ to be 
reached at some point $0< x^{*}< 1$. Notice the {\em peak structure} 
that appears in $J_{y}(x)$ and the corresponding sign change in the 
$J_z$ component. The point of vanishing $J_z$ determines the maximum 
of the peak structure in $J_y$, i.e: the full current flow is 
parallel to the Y axis.
%
%
\subsubsection{Intermediate field hysteresis}
\label{intermediatefield}
When the maximum applied field $H_{m}$ is allowed to increase beyond 
$H_{m}^{*}$ new phenomena appear. Fig.\ref{fig3} displays 
some outstanding features, which are sketched here: (i) it is 
apparent that the $H_{y}(x)$ curves do not show a conventional 
hysteresis cycle structure. Notice that the profiles no.1 and no.11 that
correspond to $H_{yS}=H_m$ do not fit. (ii) Contrary to the case of 
low field cycles, the profiles $H_{z}(x)$ do not repeat values for 
the ascending/descending branches. When the sample centre is reached 
(profile no.4) irreversible flux entry happens (no.5 and no.6). 
Further changes (curves no.7 to no.11) occur over a reduced shielding 
curve.
Finally, one can see that the aforementioned features are clearly 
translated to the current density vector $\vec{J}(x)$.

%
\subsubsection{High field hysteresis}
\label{highfield}

In this part we show that, within the isotropic model, a transverse field $H_{yS}$ may effectively collapse the 
longitudinal diamagnetic profile $H_{z}(x)$. Fig.\ref{fig4} displays 
a magnetization process in which $H_{yS}$ is increased up to the 
value $H_{m}=3$ in our dimensionless units. Then, $H_{yS}$ is 
decreased down to $H_{m}=-3$. Further increase to $H_{m}=3$ is not 
shown just to avoid intricacy in the graph. However, the complete 
hysteresis cycle will be presented in the next section. Here one can 
see that $H_{y}(x)$ follows a typical CSM high field evolution, 
whereas $H_{z}(x)$ irreversibly evolves to a flat profile, compatible 
with the boundary condition $H_{z}(1)=H_{zS}$. Simultaneously, the 
current density related to $H_{z}$ becomes zero. Thus, a constant 
trapped field $H_{zS}$ is generated within the sample. Any subsequent 
change in $H_{yS}$ will not affect the trapped field and the 
associated current density component.

The described behavior can be well understood in terms of the 
minimization principle. The process tends to minimize
\[
{\cal C}
=\frac{1}{2}\int_{\Omega}\! (H_{\rm n+1,y}-H_{\rm n,y})^{2}
+
(H_{\rm n+1,z}-H_{\rm n,z})^{2}
  \; .
\]
It is apparent that if the condition $H_{\rm n+1,z}(x)=H_{\rm 
n,z}(x)$ is fulfilled, ${\cal C}$ cannot be smaller than the minimum 
value of $1/2\int\! (H_{\rm n+1,y}-H_{\rm n,y})^{2}$, which appears 
for the full current flow dedicated to $H_y$. Thus, one has 
$J_{z}=\pm J_{c}$ and $J_{y}=0$, and both conditions hold thereafter.


\subsection{Double critical state model}
\label{dcsm}
Below we will show that our approach reproduces the results derived 
by the standard DCSM when the appropriate control space for the 
current density is chosen. For this purpose we focus on 
Ref.\onlinecite{clemgonzalez3} in which the DCSM was used for 
investigating a wealth of phenomena in rotating field experiments. In 
order to ease comparison, we shall adopt the notation therein. An 
infinite slab with surfaces at $x=0$ and $x=2X_{m}$ was considered 
with the magnetic field contained in the YZ plane.

The starting point will be the restatement of the functional ${\cal 
C}$ in appropriate generalized coordinates
\[
{\cal C}[\vec{H}_{\rm n+1}(\vec{x})]
=\frac{1}{2}\int_{\Omega}\! H_{\rm n+1}^{2}
-2H_{\rm n}H_{\rm n+1}\cos{(\alpha_{\rm n+1}-\alpha_{\rm n})} \; .
\]
Here $H,\alpha$ stand for the field modulus and the angle with 
respect to the Y axis. $H_{\rm n},\alpha_{\rm n}$ correspond to the 
given previous profile and one must solve for $H_{\rm 
n+1},\alpha_{\rm n+1}$. Now, ${\cal C}$ must be minimized with the 
constraints
\[
|\frac{dH_{\rm n+1}}{dx}|\leq J_{c\perp} \quad ; \quad
|\frac{d\alpha_{\rm n+1}}{dx}|\leq k_{c\parallel}\; ,
\]
where $J_{c\perp}$ and $k_{c\parallel}$ are the constants characterizing the 
flux pinning and cutting thresholds. Following 
Ref.\onlinecite{clemgonzalez3} $k_{c\parallel}$ relates to the parallel 
critical current density by $k_{c\parallel}=J_{c\parallel}/H$.

On defining the characteristic length scale $\lambda\equiv 
1/k_{c\parallel}$, and using $H$ in units of $J_{c\perp}\lambda$ and $x$ in 
units of $\lambda$, one obtains the dimensionless expressions
\begin{eqnarray}
\frac{dH_{\rm n+1}}{dx}&=&\frac{J_{\perp}}{J_{c\perp}}\equiv
u_{h}\quad ; \quad |u_{h}|\leq 1
\nonumber
\\[1ex]
\frac{d\alpha_{\rm n+1}}{dx}&=&\frac{k_{\parallel}}{k_{c\parallel}}\equiv
u_{\alpha}\quad ; \quad |u_{\alpha}|\leq 1
\nonumber
\end{eqnarray}
In other words, the control set $\Delta$ is a square in this units.

Next, we introduce the associated Hamiltonian 
[Eq.(\ref{eqnhamiltonian}) in Sec.\ref{hameq}]
\[
{\cal H}=p_{h}u_{h}+p_{\alpha}u_{\alpha}-
\frac{1}{2}[H_{\rm n+1}^{2}-2H_{\rm n}H_{\rm n+1}
\cos{(\alpha_{\rm n+1}-\alpha_{\rm n})}]
\; .
\]
Then, if one applies the Pontryagin's maximum principle
\[
\max_{\vec{u}\in \Delta}{\cal H}
\equiv
\max_{\vec{u}\in \Delta}(p_{h}u_{h}+p_{\alpha}u_{\alpha})
\]
$(u_{h}^{*},u_{\alpha}^{*})$ is determined again as a vector leaning on the boundary of $\Delta$. This has been illustrated in Fig.\ref{fig5} and relies on the fact that $\vec{p}\cdot\vec{u}$ is maximum for the 
largest projection of $\vec{u}$ on $\vec{p}$. If $(p_{h},p_{\alpha})$ 
are non-vanishing, one gets $(u_{h}^{*},u_{\alpha}^{*})=[{\rm 
sgn}(p_{h}),{\rm sgn}(p_{\alpha})]$, which can be identified as a 
CT-zone within the DCSM framework. On the other hand, $p_{h}\neq 
0,p_{\alpha}=0$ in an open interval will give place to a T-zone in 
which $u_{h}^{*}={\rm sgn}(p_{h})$ and $u_{\alpha}^{*}$ must be determined on the basis of other arguments. Finally, $p_{h}=0,p_{\alpha}\neq 0$ corresponds to a C-zone, in which $u_{\alpha}^{*}={\rm sgn}(p_{\alpha})$ and $u_{h^{*}}$ is yet undetermined.

Thus, in the case under consideration (i.e.: field independent 
$J_{c\perp},k_{c\parallel}$) the Hamiltonian equations are
\begin{mathletters}
\label{eqndcsm:all}
\begin{eqnarray}
\frac{dH_{\rm n+1}}{dx}&=&{\rm sgn}(p_{h})
\quad {\rm or} \quad {\rm undetermined}
\label{eqndcsm:a}
\\[1ex]
\frac{d\alpha_{\rm n+1}}{dx}&=&{\rm sgn}(p_{\alpha})
\quad {\rm or} \quad {\rm undetermined}
\label{eqndcsm:b}
\\[1ex]
\frac{dp_{h}}{dx}&=&H_{\rm n+1}-H_{\rm n}\cos{(\alpha_{\rm 
n+1}-\alpha_{\rm n})}
\label{eqndcsm:c}
\\[1ex]
\frac{dp_{\alpha}}{dx}&=&H_{\rm n+1}H_{\rm n}
\sin{(\alpha_{\rm n+1}-\alpha_{\rm n})}.
\label{eqndcsm:d}
\end{eqnarray}
\end{mathletters}
These equations can be straightforwardly solved to produce the field 
penetration profiles. Different expressions arise related to the 
mentioned zone structure.
%
%
\subsubsection{CT zone}
\label{ctzone}
If one has $p_{h}\neq 0,p_{\alpha}\neq 0$, the field penetration is 
given by the differential equations
\begin{eqnarray}
\frac{dH_{\rm n+1}}{dx}&=&{\rm sgn}(p_{h})
\quad \Rightarrow \quad H_{\rm n+1}={\rm sgn}(p_{h})x+H_{\rm n+1}(0)
\nonumber
\\[1ex]
\frac{d\alpha_{\rm n+1}}{dx}&=&{\rm sgn}(p_{\alpha})
\quad \Rightarrow \quad \alpha_{\rm n+1}={\rm 
sgn}(p_{h})x+\alpha_{\rm n+1}(0) \; .
\nonumber
\end{eqnarray}
Above, we have incorporated surface boundary conditions, just for definiteness.
%
%
\subsubsection{T zone}
\label{tzone}
In this case one has
\[
H_{\rm n+1}={\rm sgn}(p_{h})x+H_{\rm n+1}(0)
\]
and $\alpha_{\rm n+1}(x)$ is determined by the condition 
$dp_{\alpha}/dx=0$, which guarantees that $p_{\alpha}$ keeps the zero 
value in this region. Then
\[
H_{\rm n+1}(x)H_{\rm n}(x)\sin{[\alpha_{\rm n+1}(x)-\alpha_{\rm n}(x)]}=0
\]

%
\subsubsection{C zone}
\label{czone}
In this case one has
\[
\alpha_{\rm n+1}={\rm sgn}(p_{h})x+\alpha_{\rm n+1}(0)
\]
and $H_{\rm n+1}(x)$ is determined by the condition $dp_{h}/dx=0$, which gives
\[
H_{\rm n+1}(x)=H_{\rm n}(x)\cos{[\alpha_{\rm n+1}(x)-\alpha_{\rm n}(x)]}
\]

%
\subsubsection{Example: nonmagnetic initial state}
\label{example}
Several initial magnetic configurations were examined in 
Ref.\onlinecite{clemgonzalez3} within the conventional DCSM formulation. For illustration, the nonmagnetic initial state, which corresponds to the so-called field cooled experiment is analyzed here in the framework of our theory.

We assume initial uniform flux density within the superconductor 
$H_{\rm 0}(x)=H_{s},\alpha_{\rm 0}(x)=0$. Then, consider the 
quasisteady evolution towards the state $H_{\rm n+1}(x),\alpha_{\rm 
n+1}(x)$ with boundary conditions at the surface $H_{\rm 
n+1}(0)=H_{s}$ and $\alpha_{\rm 
n+1}(0)=(n+1)\delta\alpha\equiv\alpha_{s,n+1}$. The minimization of 
${\cal C}$ for the $n+1$ time layer requires an initial CT-zone given by
\begin{eqnarray}
H_{\rm n+1}(x)&=&H_{s}-x\quad ; \quad p_{h}<0
\nonumber
\\[1ex]
\alpha_{\rm n+1}(x)&=&\alpha_{s,n+1}-x\quad ; \quad p_{\alpha}<0 \; .
\nonumber
\end{eqnarray}
The CT-zone spreads up to the point $x_{v,n+1}$ where the 
conditions $p_{h}=0,dp_{h}/dx=0$ are reached. This point may be determined by iteration of the governing Hamiltonian equations (Eqs. (\ref{eqndcsm:all})) over the previous steps. Such process leads to
\begin{eqnarray}
H_{s}-x_{v,n+1}-H_{\rm n}(x_{v,n+1})\cos{(\delta\alpha)}=0 \qquad \Rightarrow
\nonumber
\\[1ex]
x_{v,n+1}=H_{s}\left[ 1-\cos{(\alpha_{s,n+1}-x_{v,n+1})}\right]
\nonumber
\end{eqnarray}
and defines the beginning of a C-zone. The equation determining $x_{v,n+1}$ has been obtained within the approximation $\delta\alpha\to 0\Rightarrow\cos^{n}{(\delta\alpha)}\to 1$. This condition yields the continuum limit of our discretized approach. Notice that our expression matches the one obtained by the conventional DCSM.
Finally, the C-zone stretches from $x_{v,n+1}$ to $x_{c,n+1}$, the point where $\alpha_{\rm n+1}$ vanishes. Then, the field modulus profile becomes
\begin{equation}
H_{\rm n+1}(x)=H_{s}
\cos{(\alpha_{s,n+1}-x)} \; .
\end{equation}
The associated momenta $p_{h},p_{\alpha}$ may be straightforwardly 
obtained by integration of their first order differential equations. In Fig.\ref{fig6} we have plotted the first step profiles $H_{\rm 1}(x),\alpha_{\rm 1}(x),p_{h}(x),p_{\alpha}(x)$. Several points have been marked on the curves $p_{h}(x)$ and $p_{\alpha}(x)$, that correspond to the sequence of values of the control variable $\vec{u}$ standing out in Fig.\ref{fig5}. Notice that $(u_{h},u_{\alpha})$ always takes values on the boundary of the control space. Within a CT-zone $\vec{u}$ remains on a vertex of the square. As soon as a component of 
$\vec{p}$ vanishes (point 3 in our example) the associated component 
of $\vec{u}$ must be determined by the condition of zero derivative. 
In our case, this produces a jump to the vector $\vec{u}_{3}$, which 
defines the beginning of the C-zone 
$(\vec{u}_{3},\vec{u}_{4},\vec{u}_{5},\dots)$. This jump in the control variable is known as {\em bang-bang} phenomenon among mathematicians. Eventually, if the center of the sample is not reached by the variations, a new jump in the current density gives way to the O-zone ($u_{h}=0,u_{\alpha}=0$).
%
%
%
\section{magnetization loops}
\label{magloops}
Though some experiments have been devised for the direct measurement 
of the field or current penetration profiles, the common practice is 
to record average values. Thus, the vast majority of data on vortex 
matter in type II superconductors consist of magnetic moment 
measurements. Recall that the macroscopic magnetization of the sample 
is defined as $\vec{M}\equiv\langle \vec{H}(x)\rangle -\vec{H}_{S}$, 
where $\vec{H}_{S}$ stands for the uniform applied field.

In this section we concentrate on the magnetization loops which can 
be derived from the previously treated field penetration profiles. It 
will be shown that important experimental results can be reproduced. 
New observable features are also predicted, which may be used as a 
test for the physically meaningful critical current control space. 
All the results presented below correspond to the isotropic 
hypothesis, which has been mainly developed in this paper. The 
comparison to the already well established DCSM and a discussion on 
the selection of the control space for the current density are 
considered in the next section.

First, we analyze the influence of the maximum applied field $H_m$ on 
the magnetization loop. Fig.\ref{fig7} collects the results 
corresponding to the profiles analyzed in SubSec.\ref{circ}: low 
field ($H_{m}<H_{m}^{*}$), intermediate field ($H_{m}>H_{m}^{*}$), 
and high field ($H_{m}\gg H_{m}^{*}$). We have depicted both $M_y$ 
and $M_z$ for the zero field cooled process: (i) Apply $H_{zS}$, (ii) 
increase $H_{yS}$ from zero to $H_{m}$, (iii) decrease $H_{yS}$ to 
$-H_{m}$, and (iv) increase $H_{yS}$ to $H_{m}$ again.

$M_{y}(H_{yS})$ roughly displays a standard behavior, but shows a 
noticeable peculiarity for the intermediate field loop. The loop 
fails to close after a complete field cycle. This feature was not 
noticed for the low/high field loops, at least to the numerical 
precision of our calculations. The $M_{z}(H_{yS})$ curves show a 
quite less conventional behavior as compared to one dimensional 
measurements in which a single field component appears. In addition, qualitatively different tendencies can be observed in terms of the 
value $H_m$. Thus, low field cycles produce a {\em butterfly} loop; 
intermediate fields produce a descending behavior for the transverse 
magnetic moment, and high fields produce the irreversible collapse of 
magnetization.

As a final observation, we describe an outstanding feature that was 
observed in our simulations. A second cycle in the applied field 
$H_{yS}$ succeeds to close the hysteresis loop $M_{y}(H_{yS})$ in the 
intermediate field region. Meanwhile, $M_{z}$ keeps the descending 
trend, though seeming to stabilize around a non-vanishing value. This 
has been depicted in Fig.\ref{fig8}.

We want to emphasize that our intermediate field simulations 
reproduce the experimental observations in Ref.\onlinecite{park} and 
in some related papers cited therein. Thus, the isotropic 
hypothesis sets a simple model for those crossed field measurements. 
Furthermore, we suggest to extend such measurements as a test of the 
theory, i.e.: perform low/high field loops and a second cycle in 
order to check the predicted features.
%
%
\section{Discussion and conclusions}
\label{conclusions}
We have introduced a variational formulation that generalizes Bean's 
critical state model for hard superconductors, while keeping its conceptual simplicity. In the spirit of the original theory, we have also avoided the explicit use of the electric field $\vec{E}$. However, the time evolution of the magnetic field profiles $\vec{H}(x,t)$ can be obtained by replacing $\vec{E}$ with 
an appropriate restriction on the current density $\vec{J}$ flowing 
within the sample. Mathematically, a wide range of possibilities are 
allowed as we pose it in the form $\vec{J}\in\Delta (\vec{H},x)$ with 
$\Delta$ a bounded set. Upon the selection of a definite set 
$\Delta$, our theory may be considered as the background for a variety 
of critical state models. In particular, we have analyzed 
the cases of choosing either a circle or a rectangle (isotropic model 
and DCSM respectively).

The variational quantity which is minimized under the aforementioned 
restriction has got a clear physical interpretation and quite general 
validity. We propose that the time evolution is governed by the 
balance between a screening term (inertia) and an entropy production 
term. This relies on basic principles of non-equilibrium 
thermodynamics.\cite{prigogine} Within the usual assumption of 
instant response (vertical $\vec{E}(\vec{J})$ graph) the 
singular behavior of $\dot{\cal S}=\vec{E}\cdot\vec{J}/T$ is treated 
by replacing the associated term with the condition $\vec{J}\in\Delta$ in the minimization principle. This kind of statement fits the mathematical theory of optimal control,\cite{oc} that has been applied in this paper. The method is appropriate for the numerical treatment of realistic problems in which the magnetization is a vector quantity. Such situation occurs in experimental setups designed for studying the interaction of twisted vortices and in the field of technological applications of hard type-II superconductors.

We want to emphasize that the selection of the permitted set $\Delta$ 
for the current density should be either justified in terms of 
fundamental theories or bounded by experimental observations. The 
scope of our paper is to provide a suitable mathematical treatment 
for general CSM problems and a physical background for the method. A 
well posed inversion theory, allowing to obtain the set $\Delta$ from 
experiments is still an open question. 

However, some conclusions can 
be drawn from the comparison of simulations with different sets 
$\Delta$. On the one side, some experimental observations are quite independent of the region $\Delta$ in use. This is the case of the field penetration profiles in rotating field experiments. A very similar behavior is predicted either by the isotropic or DCSM hypothesis.\cite{badia2} On the other side, one can find outstanding differences between these two models. Firstly, the former predicts the criticality condition $J=J_c$ for those regions which have been affected by an electric field. Within the DCSM framework, however, 
one can obtain {\em critical} regions (CT zones) where 
$J=J_c=\surd (J_{c\parallel}^ {2}+J_{c\perp}^{2})$ and {\em subcritical} 
regions (C or T zones) where $J<J_c$. Microscopic probes can be used 
to investigate this property. For instance, Ref.\onlinecite{gordeev} 
reports neutron polarization analysis of the magnetic field profiles 
within ceramic YBCO samples subjected to rotation stages. The 
isotropic hypothesis seems to be favoured for this Josephson medium. 
On the macroscopic side, one can design adequate experiments which 
check the compatibility with specific selections of $\Delta$. The 
investigation of the so-called {\em magnetization 
collapse}\cite{voloshin} is a promising possibility. In fact, recent experimental results on melt-textured and single crystal YBCO samples\cite{fisher1,fisher2} are nicely described within the isotropic model. The authors of Ref.\onlinecite{fisher2} have studied the transverse magnetic moment $M_z$ suppression by cycling the longitudinal magnetic field $H_{yS}$. As the most important observed feature they emphasize the symmetric decrease of $M_z$ for the diamagnetic and paramagnetic initial states of the sample. This is shown to be in clear contradiction with the DCSM hypothesis and qualitatively interpreted by their {\em two-velocity hydrodynamic model}. In Fig.\ref{fig9} we show that our general critical state theory gives a quite satisfactory explanation of the experimental facts. We get a symmetric suppression of $M_z$. The field penetration profiles show that the slope decrease in $H_z(x)$ near the surface, due to the enhancement of the current density component which shields the change in $H_y$, occurs in a quite similar fashion, regardless the initial magnetic state. Additionally, we want to notice that the behavior of $M_z$ upon a second cycle in $H_{yS}$ which is observed in our simulations (see Fig.\ref{fig8}) gives further evidence of the validity of the model. The coincidence with the experimental data depicted in Fig.5 of Ref.\onlinecite{fisher2} is remarkable.

Finally, we will comment on some future extensions of our work. Here 
we have used a time-discretization scheme, and a slab geometry, in 
order to produce an ordinary differential equation statement of our 
principle. This simplifies the theory, but should not be considered 
as a limitation. Optimal control for systems governed by partial 
differential equations may be applied for more general cases. Thus, 
one could afford a continuum approach to the problem and also include 
the sample finite size effects if necessary. Other issues as the use 
of different physically meaningful control sets $\Delta$ (f.i.: 
elliptic, so as to incorporate material anisotropy), spatial inhomogeneity or surface barrier effects are also suggested.
%
%
\section*{acknowledgements}
The authors acknowledge financial support from Spanish CICYT (projects 
MAT99-1028 and  BFM-2000-1066/C0301).
%
%
%
%

%
%
%
%
%
\begin{figure}
\caption[initial curves]{
Field and current density penetration profiles for a zero field 
cooled slab. The dots indicate the initial process in which $H_z$ was 
raised to the value $H_{zS}=0.6$ at the surface. The successive steps 
increasing $H_{yS}$ are labelled from 1 to 5. Continuous lines are 
used for $H_{z}(x)$ and the associated current density $-J_{y}(x)$, 
while dashed curves represent $H_{y}(x)$ and $J_{z}(x)$. 
Dimensionless units are used as defined in the text.
\label{fig1}}
\end{figure}
\vspace{1cm}
\begin{figure}
\caption[small amplitude]{
Field and current density penetration profiles for a small amplitude 
cycle in $H_{yS}$, successive to the initial magnetization process 
described in Fig.\ref{fig1}. The descending branch curves are 
labelled 1 to 7 (continuous lines), whereas ascending curves are 
labelled 8 to 14 (dash-dotted curves). The $H_{z}(x)$ and $-J_{y}(x)$ 
profiles nearly repeat values for the descending/ascending processes 
as one can see in the graph. Some labels have been skipped to avoid 
confusion.
\label{fig2}}
\end{figure}
\vspace{1cm}
\begin{figure}
\caption[medium amplitude]{
Field and current density penetration profiles for an intermediate 
amplitude cycle in $H_{yS}$, successive to the initial magnetization 
process described in Fig.\ref{fig1}. The curves labelled 1 to 6 
(continuous lines) stand for the descending process, while the curves 
7 to 11 (dash-dotted) stand for the ascending part. The label no.7 in 
the family of curves $H_{z}(x)$ has been avoided for clarity.
\label{fig3}}
\end{figure}
\vspace{1cm}
\begin{figure}
\caption[high amplitude]{
Field and current density penetration profiles corresponding to a 
high amplitude cycle in $H_{yS}$, successive to the initial 
magnetization process described in Fig.\ref{fig1}. Here, we have 
labelled 1 to 7 the profiles for the initial increase to the high 
amplitude ($H_{yS}=3$). The descending branch curves are labelled 8 
to 14. In order to avoid confusion, the family of ascending branch 
curves of the cycle are not displayed.
\label{fig4}}
\end{figure}
\vspace{1cm}
\begin{figure}
\caption[control space dcsm]{
Diagram of the square control space $\Delta$ on purpose for the DCSM 
oriented according to the local magnetic field. For illustration, we 
plot the optimal control solution $\vec{u}$ which maximizes 
$\vec{p}\cdot\vec{u}$ in a generic situation. Some specific values 
$\vec{p}_{1},\vec{p}_{2},\vec{p}_{3},\dots$ and the corresponding 
solutions $\vec{u}_{1},\vec{u}_{2},\vec{u}_{3},\dots$ are also 
marked. The sequence is extracted from the example depicted in 
Fig.\ref{fig6}.
\label{fig5}}
\end{figure}
\vspace{1cm}
\begin{figure}
\caption[example dcsm]{
Normalized field modulus and angle penetration profiles 
($H(x)/H_{S},\alpha (x)/\alpha_{S}$) obtained by the application of 
our variational principle to the DCSM control space. A field cooled slab was assumed, subjected to further rotation of the surface field. Also 
depicted are the associated momenta ($p_{h}(x),p_{\alpha}(x)$). The 
points 1 to 5 labelled on the momenta curves correspond to the 
sequence of control values shown in Fig.\ref{fig5}. Following the 
notation introduced in Ref.\onlinecite{clemgonzalez3}, we have marked 
the CT, C and O zones, as well as the penetration points $x_v$ and 
$x_c$. The distance $x$ is given in units of the slab half-width 
$x_m$. Contrary to our choice in the rest of figures, in this case 
$x=0$ corresponds to the surface of the slab.
\label{fig6}}
\end{figure}
\vspace{1cm}
\begin{figure}
\caption[magnetization loops]{
Evolution of the magnetization components in a simulated crossed 
field experiment for a zero field cooled superconducting slab. 
Subsequent to the application of a constant surface field $H_{zS}$, 
the other component $H_{yS}$ was cycled ($0\rightarrow H_m 
\rightarrow -H_m \rightarrow H_m$) for three different values of 
$H_m$. Continuous lines stand for low fields, continuous-dotted for 
intermediate fields and dashed lines represent the high field cycle.
\label{fig7}}
\end{figure}
\vspace{1cm}
\begin{figure}
\caption[second cycle]{
First and second magnetization cycles for the intermediate field loop 
considered in Fig.\ref{fig7}.
\label{fig8}}
\end{figure}
\vspace{1cm}
\begin{figure}
\caption[collapse]{
Dependence of the transverse magnetization $M_z$ on the first cycle of $H_{yS}$ ($0\rightarrow H_m \rightarrow -H_m \rightarrow H_m$) for the para- and diamagnetic initial states in $H_{z}(x)$. The upper panel shows the suppresion of $M_z$ in the paramagnetic case (sequence O'A'B'C') and the lower panel reflects the same phenomenon for the diamagnetic case (sequence OABC). The insets illustrate some selected penetration profiles of $H_{z}(x)$, corresponding to the points labelled on the $M_z$ curves.
\label{fig9}}
\end{figure}

\end{document}